\documentclass[journal]{IEEEtran}
\usepackage{cite}
\ifCLASSINFOpdf
   \usepackage[pdftex]{graphicx}
\else
\fi
\usepackage{amsmath}
\usepackage{amssymb}
\usepackage{xcolor}
\begin{document}
\title{Wasserstein GANs for MR Imaging:\\ from Paired to Unpaired Training\vspace{5mm}}

\author{\large{Ke~Lei$^1$,~\IEEEmembership{}
        Morteza Mardani$^{1,2}$,~\IEEEmembership{}
        John M. Pauly$^1$,~\IEEEmembership{}
        and Shreyas S. Vasanawala$^2$~\IEEEmembership{}\vspace{3mm}% <-this % stops a space
        
 Departments of $^1$Electrical Engineering, $^2$Radiology.\  Stanford University.}\\
%Emails: \texttt{kllei}, \texttt{morteza}, \texttt{pauly}, \texttt{vasanawala}\texttt{@stanford.edu}.

\thanks{$\dag$ Work in this paper was supported by the NIH R01EB009690 and NIH R01EB026136 award, and GE Precision Healthcare. Part of the results have been submitted to and presented at the $27$th annual meeting of International Society of Magnetic Resonance in Medicine (ISMRM), Montreal, Canada, May 2019.}% <-this % stops a space
\thanks{Copyright (c) 2019 IEEE. Personal use of this material is permitted. However, permission to use this material for any other purposes must be obtained from the IEEE by sending a request to pubs-permissions@ieee.org.}} 
%\thanks{J. Doe and J. Doe are with Stanford University.}% <-this % stops a space
%\thanks{Manuscript received April 19, 2019; revised August 26, 2015.}}

\bstctlcite{IEEEexample:BSTcontrol}

\markboth{IEEE TRANSACTIONS ON MEDICAL IMAGING}{ }

\maketitle

\begin{abstract}
Lack of ground-truth MR images impedes the common supervised training of neural networks for image reconstruction. To cope with this challenge, this paper leverages \textit{unpaired} adversarial training for reconstruction networks, where the inputs are undersampled k-space and naively reconstructed images from one dataset, and the labels are high-quality images from another dataset. The reconstruction networks consist of a generator which suppresses the input image artifacts, and a discriminator using a pool of (unpaired) labels to adjust the reconstruction quality. The generator is an unrolled neural network -- a cascade of convolutional and data consistency layers. The discriminator is also a multilayer CNN that plays the role of a critic scoring the quality of reconstructed images based on the Wasserstein distance. Our experiments with knee MRI datasets demonstrate that the proposed unpaired training enables diagnostic-quality reconstruction when high-quality image labels are not available for the input types of interest, or when the amount of labels is small. In addition, our adversarial training scheme can achieve better image quality (as rated by expert radiologists) compared with the paired training schemes with pixel-wise loss. 
\end{abstract}


% \fi
\IEEEpeerreviewmaketitle

\section{Introduction}

\IEEEPARstart{M}{agnetic} resonance imaging (MRI) is commonly used clinically for its flexible contrast. The major shortcoming of MRI is its long scan time, especially for volumetric images. Undersampling is often necessary to reduce scan time and cope with motion, but reconstructing undersampled MRI is solving an undetermined system and conventional reconstruction methods such as compressed sensing (CS) are time intensive. Recently, data-driven methods based on neural networks (NNs) are adopted to reconstruct MR images with rapid reconstruction speed. However, most of these models require supervised training on a large and specific set of labels, that are fully-sampled high-quality images. We refer to the label image used for training supervision as `label' in this paper.

Collecting such labels is expensive or impossible in certain scenarios such as dynamic imaging. For instance, in dynamic contrast enhanced (DCE) imaging, the contrast is rapidly changing, or, for deformable moving organs in the chest, abdomen, or pelvis with respiratory motion, acquiring the ground truth image is a daunting task. On the other hand, basic 2D scans for static body parts, such as extremities and brain, are often fully-sampled with high quality to serve as labels. We aim to train a model for cases where there are no, or, only limited ground truth images. This is possible with unpaired training, where the labels can be different from the images being reconstructed (i.e. the inputs).

Ample research has been conducted during the last few years on deep learning for MRI reconstruction \cite{nature,2019arXiv190401112K,doi:10.1002/jmri.26871,Hyun2018,2019arXiv190307824C,Qin2017ConvolutionalRN,8067520}. The majority of those works use paired training which demands a large number of labels specifically for the task they are tackling. There are only a few attempts to cope with label scarcity, as in \cite{Tamir,feiyuabs,abs-1901-04547,abs-1710-02615,N2N}, using self-supervision and transfer learning. Unpaired training with adversarial objectives is an alternative that has been explored in computer vision for natural image translation tasks \cite{CycleGAN2017}. However, for medical imaging tasks, it introduces the risk of hallucinating images that may adversely affect the subsequent diagnosis. The methods in~\cite{abs-1812-11440,DAGAN,GANCS,cycMRI,abs-1902-06455,SR}, although adopting adversarial objectives, are still paired and rely heavily on some pixel-wise supervision, such as the $\ell_1$ distance, for stabilizing the training and reducing the hallucination risk. Adversarial methods used in these works were adopted from entropic generative adversarial networks (EGANs) \cite{GAN} or least-squares GANs (LSGANs) \cite{lsgan}. Without the pixel-wise supervision, these methods return images with coherent artifacts. Aside from deep learning approaches, there are blind learning based techniques \cite{soup,strollr} that exploit low-rank and sparsity regulation and do not require label image. However, these techniques are much slower than even CS.

\begin{figure}  [!h]
\begin{center}
	\includegraphics[scale = 0.13] {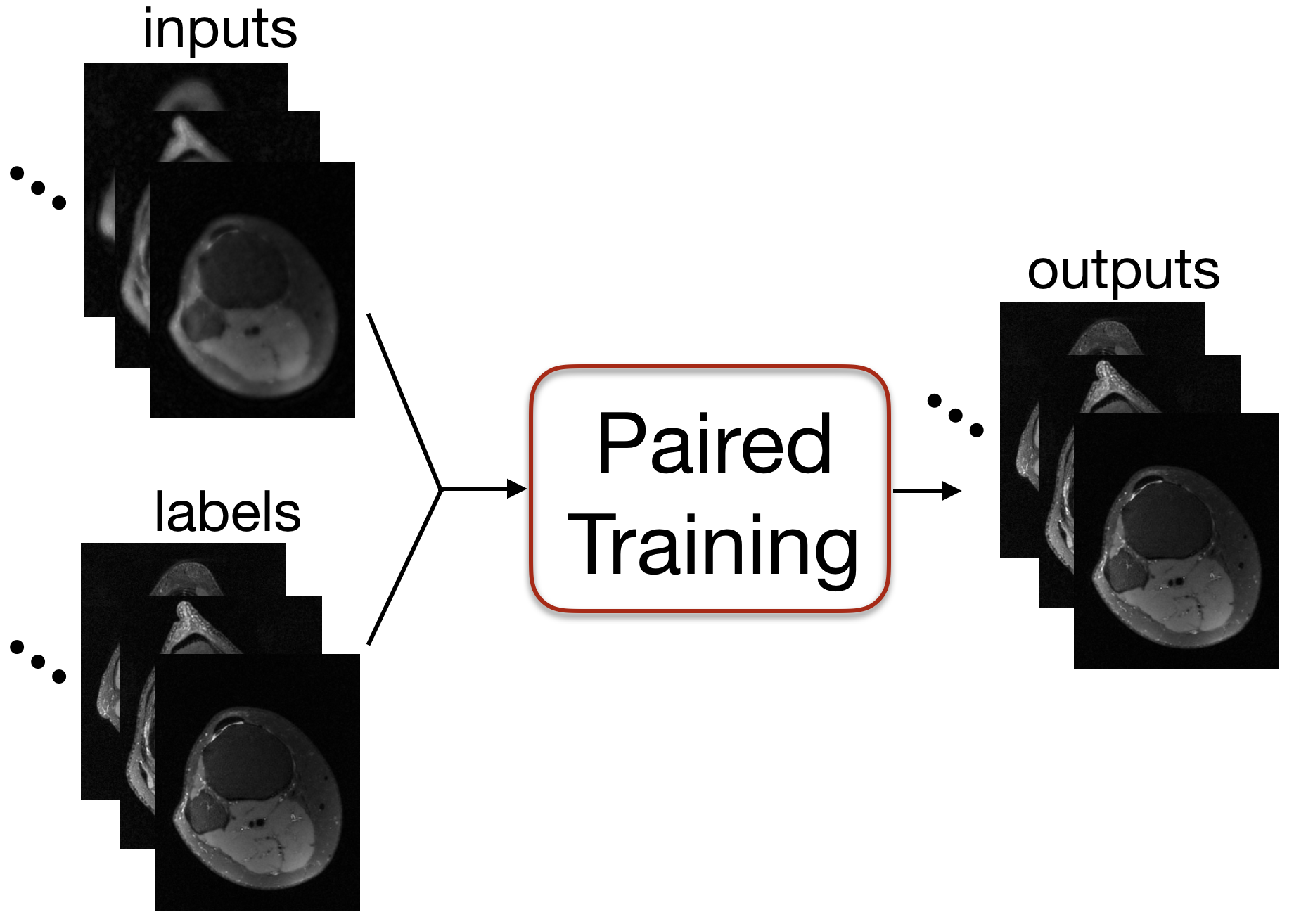}\hspace{2.7mm}
	\includegraphics[scale = 0.13] {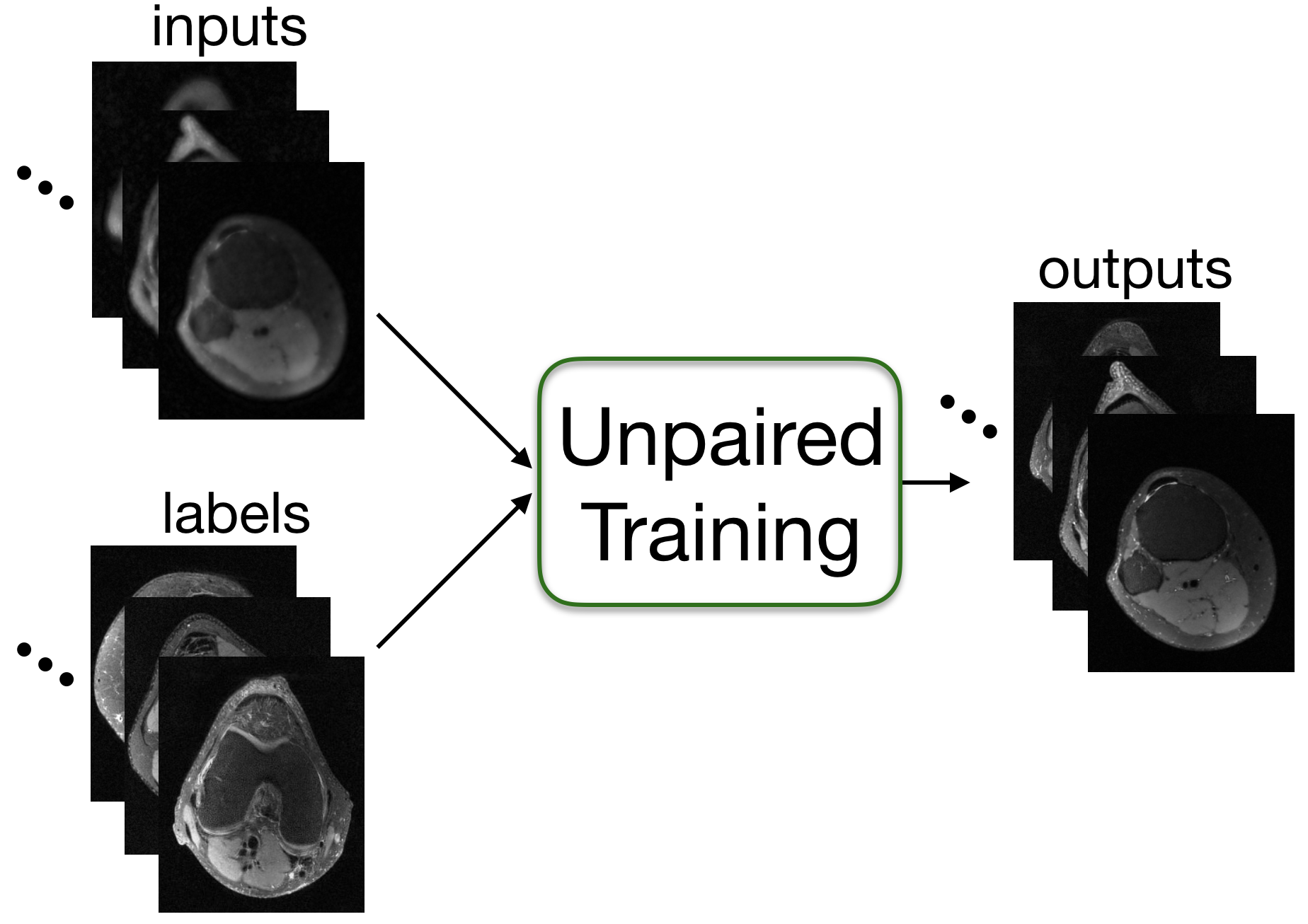}
    \caption{A high-level illustration of paired (left) vs. unpaired (right) training.}
    \label{ipunp}
\end{center}
\end{figure}
\vspace{-0.2cm}
 We introduce an unpaired training scheme for MRI reconstruction by leveraging adversarial training based on the Wasserstein distance \cite{WGAN} and data consistency (DC). Our training scheme involves two networks, a generator (G) and a discriminator (D), which are trained simultaneously and interactively. The G performs the reconstruction by taking in the undersampled k-space and outputting diagnostic-quality images. It can take various kinds of network architectures and types of data consistency. The D is a multilayer CNN that takes in the image reconstructed by the G and the image in the label set and outputs a real number that reflects the distance between the two, derived from the Wasserstein distance \cite{Villani2009}. Our model learns to approximate a desired distribution by this adversarial training process which does not require pairing between the input and label. 
 
Our proposed scheme is examined with different NN models under different scenarios of label availability. 
\textcolor{black}{We perform experiments with various settings on real-world knee and abdominal DCE MRI datasets. For the settings presented, we observe that: 1) unpaired training can be used for disjoint and partial label cases (defined in II.A); 2) Wasserstein distance based adversarial training is more suitable for unpaired training; 3) our proposed unpaired training is better than paired training using $\ell_1$-based loss; and 4) when paired training is possible, using a combination of Wasserstein distance based adversarial training and $\ell_1$-supervision training has the best overall performance of all methods examined. }    

\noindent\textit{Notation}. The operators  $\mathbb{E}[\cdot]$,  $(\cdot)^{\mathsf{H}}$, $\odot$, $\mathcal{F}\{\cdot\}$, and $\mathcal{F}^{-1}\{\cdot\}$ denote the statistical expectation, matrix Hermitian, Hadamard product, 2D discrete Fourier transform (DFT), and inverse 2D discrete Fourier transform (IDFT), respectively. $\|\cdot\|_1$ and $\|\cdot\|$ refers to the $\ell_1$-norm and $\ell_2$-norm, respectively.

\section{Problem statement and preliminaries}
\subsection{Problem statement}

MRI reconstruction, in a simplified standard setting, solves a linear inverse system $Y=\Phi(y) + u$, where $\Phi$ captures the forward model of an MRI examination and $u$ captures the noise and uncertainties in the system, to find image $y \in \mathbb{C}^n$ from partial frequency domain samples $Y \in \mathbb{C}^m$ ($m < n$). Our goal is to learn an inverse mapping $G$ so that for test data $Y$ we can automatically recover its corresponding $y$ as $G(Y)$. We approximate this mapping by a trained NN. Normally, training such a NN requires a set of inputs $I=\{Y_i\}^M_{i=1}$ and a set of 
corresponding labels $L=\{y_j\}^M_{j=1}$ because a traditional pixel-wise supervised training objective is defined on pairs of $Y_i$ and $y_j$ where $i=j$. We use \textit{paired training} in this case. 

\begin{figure}  [!h]
\begin{center}
	\includegraphics[scale = 0.24] {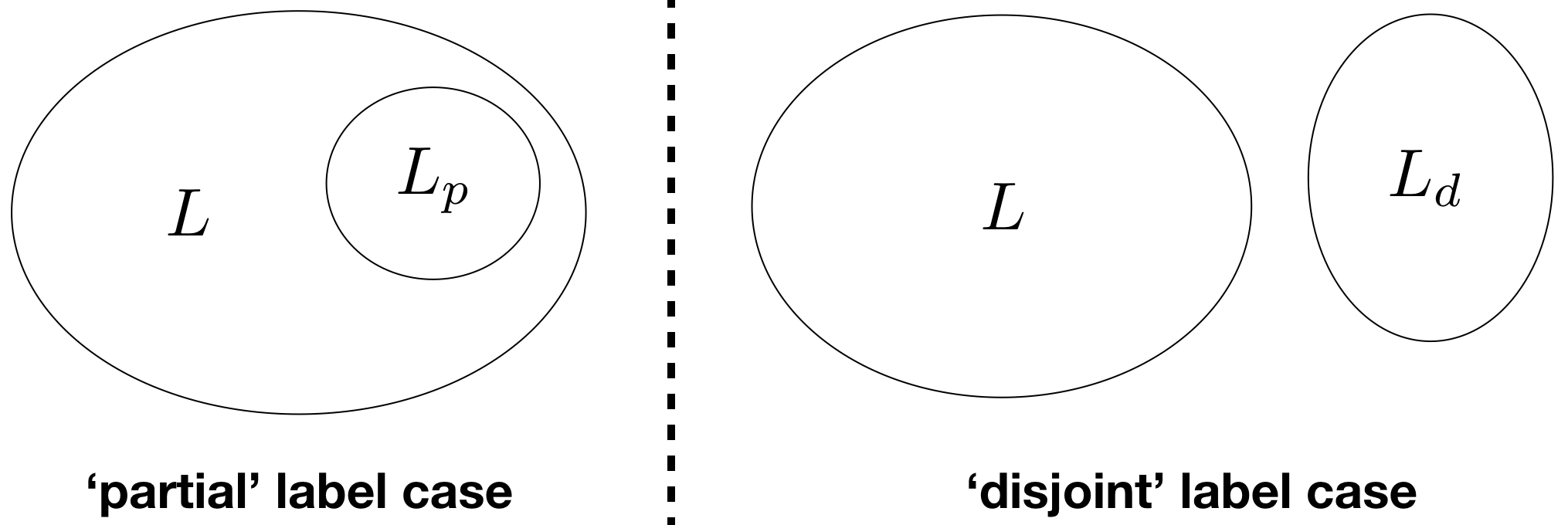}
    \caption{{Venn diagrams for two cases of label availability under \textit{unpaired} training. Given the inputs $I$, $L$ is the set of labels required for \textit{paired} training, which is unavailable. {Left: $L_p$ is a set of labels corresponding to part of the inputs. Right: $L_d$ is a set of labels disjoint with all inputs}. }}
    \label{Venn}
\end{center}
\end{figure}

{In this paper, we consider two scenarios where a set of noisy inputs $I$ is easily available but its corresponding label set $L$ is not available. First, we have a `partial' label set $$L_p=\{y_j\}^N_{j=1}\, \quad N \ll M,\ $$where $L_p\subset L$. Second, we consider a `disjoint' label set $$L_d=\big\{y_k\big\}^{\tilde{N}}_{k=1},$$where $\ L_d\cap
L=\varnothing.$ That is, for our training dataset, we either have a limited number of labels for the inputs, or, a different set of inexpensive labels. Therefore, pairs of $Y$ and $y$ cannot be used for the training. We use \textit{unpaired training} in these cases (Fig. \ref{Venn}). In the sequel, we use adversarial training based on the Wasserstein distance for unpaired learning.}

\subsection{Wasserstein distance}
Wasserstein distance is a measure of the distance between two probability distributions~\cite{Villani2009}. We particularly look at Wasserstein-1 distance in this paper. Here we first introduce Wasserstein-1 distance in its original definition that can be very computationally expensive to train with on a large set of images, then transform it into a form which can be approximated by computationally efficient training objectives. 

\begin{figure}  [!t]
\begin{center}
	\includegraphics[scale = 0.15] {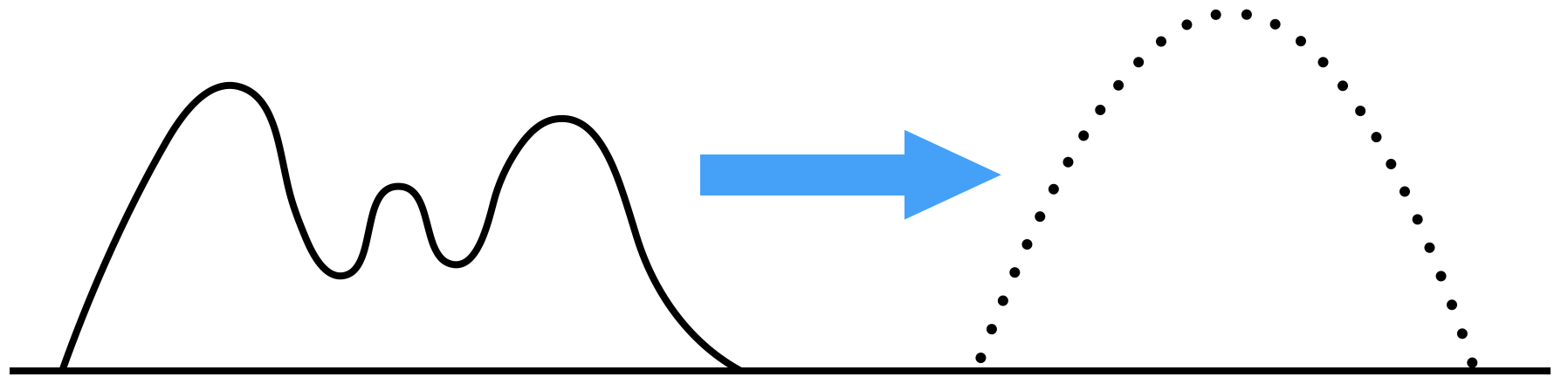}
    \caption{Illustration of the sand pile transport for earth mover's distance.}
    \label{em}
\end{center}
\end{figure}

Wasserstein-1 distance is also known as the earth-mover's (EM) distance (see Fig. \ref{em}). This quantity intuitively reflects the minimum cost (i.e., mass times distance) to transport a pile of sand to another pile (with different location and shape). One advantage of this metric is that it is continuous and differentiable almost everywhere, unlike the Jensen-Shannon (JS) divergence deployed by the original EGANs \cite{GAN}, and Pearson Chi-square divergence deployed by the LSGANs \cite{lsgan}. Generally, the Wasserstein-1 distance between probability mass $P_r$ and $P_g$ is defined as
\begin{equation}
     W(P_r, P_g)=\inf_{J\in\mathcal{J}(P_r, P_g)} \mathbb{E}_{(a,b)\sim J}\big[\hspace{0.5mm}d(a,b)\hspace{0.5mm}\big]
    \label{inf}
\end{equation} 
where $d: \mathcal{X}\times\mathcal{X}\rightarrow\mathbb{R}$ is an underlying distance metric and is chosen by convention to be the $\ell_2$ distance \cite{bwgan}, i.e., $d(a,b)=\|a-b\|$. $\mathcal{J}(P_r, P_g)$ is the set of all joint distributions for $a$ and $b$ whose marginals are $P_r$ and $P_g$ (both defined on a compact space $\mathcal{X}$), respectively.

The infimum in (\ref{inf}) is highly intractable, but using Kantorovich-Rubinstein duality \cite{Transport} one can alternatively write it as
\begin{equation}
    W(P_r, P_g)=\sup_{\|f\|_L\leq 1}\mathbb{E}_{a\sim P_r}\big[f(a)\big]- \mathbb{E}_{ b\sim P_g}\big[f(b)\big]
    \label{sup}
\end{equation}
where the supremum is over all 1-Lipschitz functions $f: \mathcal{X}\rightarrow\mathbb{R}$. $f $ is 1-Lipschitz if $\big|f(a)-f(b)\big|\leq\|a-b\|$. In practice, there are many ways to enforce or approximately enforce the 1-Lipschitz constraint. \cite{WGAN_2} introduces a computationally efficient way using gradient-norm regularization as would be discussed later. According to \cite[Proposition 1]{WGAN_2}, there is an 1-Lipschitz function $f^*$ which maximizes $\mathbb{E}_{a\sim P_r}[f(a)]- \mathbb{E}_{ b\sim P_g}[f(b)]$. This $f^*$ has gradient norm equal to $1$ almost everywhere under $P_r$ and $P_g$. So we aim to search for a $f$ whose gradient norm is close to $1$ in order to minimize the Wasserstein distance.

\section{Adversarial training of neural networks}
Theoretically, adversarial training is derived to learn a desired probability distribution by minimizing some distance between the generated data (i.e. model outputs) distribution and the label data distribution. In this paper, we use the Wasserstein-1 distance introduced in the previous section. Practically, adversarial training refers to the scheme when two networks, the generator (G) and discriminator (D), are trained simultaneously with feedback from each other's output. This scheme is used for our unpaired training approach as it does not require pixel-wise supervision. Adversarial training can also be combined with pixel-wise supervised training in the paired case.

\subsection{Unpaired training}
The ground-truth label is often not present for certain imaging scenarios. It is thus important to replace the pixel-wise supervision completely so that no pairing is needed. Then, one can leverage the available labels from other datasets that are more amenable to fully sampled acquisition. For instance, using 2D images as training labels for cine or 3D imaging inputs. In principle, adversarial training aims to approximate, in terms of some measures of distances, a probability distribution of interest: a distribution of images in the label set. Thus there is no need for each specific label to be the corresponding ground-truth of the input. Unpaired training has proved successful in image style transfer tasks (for instance, converting zebras to horses, and vice versa) such as \cite{CycleGAN2017} with adversarial training alone. These tasks in natural images do not necessarily require authentic output images. 

Adversarial training without paired supervision for medical images, however, introduces a hallucination risk. The pixel authenticity is crucial and needs to be guaranteed. Fortunately, for the considered de-aliasing problem one has the k-space data and the forward model at hand to somewhat enforce the G outputs to adhere to the k-space data. This is ensured by the DC layers embedded into the G network. DC partially alleviates the hallucination risk, but {the unstable training of GANs can still lead to reconstructions with heavy artifacts. The unstable training mainly emanates from the divergence measure from which the adversarial training objective is derived from. EGANs and LSGANs training objectives are derived from JS and Pearson Chi-square divergences, respectively. Let $P_r$ and $P_g$ denote the label and output distributions, respectively. Under an optimal D which approximates the divergence between $P_r$ and $P_g$, the gradient used to update G is the derivative of that divergence with respect to the G network parameters. When $P_r$ and $P_g$ are disjoint, JS and Chi-square divergences are not continuous, thus the gradient can become \textit{infinity}. There are also regions where this divergence is locally constant so that the gradient \textit{vanishes}. See Fig. 1 and Fig. 2 in \cite{WGAN} for examples of these two cases of uninformative gradients. The goal of generative adversarial training is to train a G where $P_g$ well approximates $P_r$. The uninformative gradients impede converging G to the minimum divergence using stochastic gradient descent (SGD) and  lead to unstable training.  }

\begin{figure}  [!t]
\begin{center}
	\includegraphics[scale = 0.39] {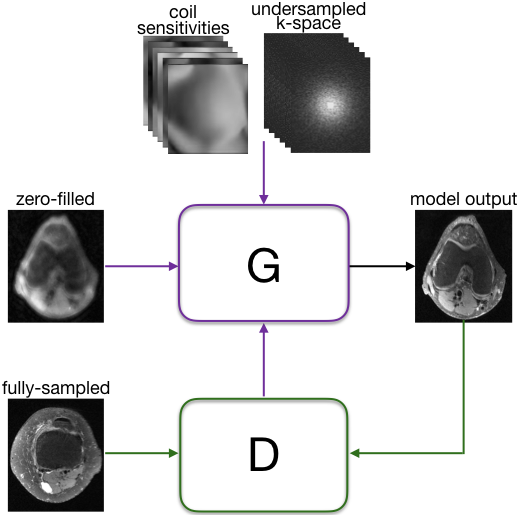}
    \caption{\small{Unpaired adversarial training.}}
    \label{flow}
\end{center}
\end{figure}

Wasserstein-1 distance is continuous even under disjoint or discrete distributions. Note that image distributions are in high-dimensional spaces and often disjoint. Therefore, we use Wasserstein GAN (WGAN) \cite{WGAN} objectives, derived from the Wasserstein-1 distance, for our unpaired training. Figure \ref{flow} illustrates the unpaired training procedure of our model. Intuitively, a D network serves as a critic which scores the images reconstructed by a G network by giving an estimate of the Wasserstein-1 distance between the G output and the label, and the G is optimized based on the feedback from D. Formally, the D network serves the role of $f$ in equation \eqref{sup} and we train it to approximate $f^*$. Since our goal is for the reconstructed images to be as good as the labels, let the labels $y\sim P_r$ and the output from G $G(x_{\rm zf})\sim P_g$. The G aims to minimize $W(P_r,P_g)$ with a given $f$. Then from equation \eqref{sup} (under some assumptions \cite{WGAN}) we have the principle version of the adversarial training objective
\begin{equation}
    \min_G\max_{\|D\|_L\leq 1}\mathbb{E}_{y\sim P_r}[D(y)]- \mathbb{E}_{G(Y)\sim P_g}[D(G(x_{\rm zf}))]
    \label{minmax}
\end{equation}
where the maximum is over all 1-Lipschitz functions $D$.

As introduced in II.B, the Lipschitz constraint on the D is enforced by searching for the 1-Lipschitz function $f^*$ which has the end-to-end gradient norm equal to unity. Considering the amount of training iterations, it is not necessary to compute and enforce this gradient norm everywhere. Therefore, \cite{WGAN_2} introduces the gradient penalty (GP) term that penalizes the gradient norm of D w.r.t. random samples drawn from real and fake distributions from diverging from 1. Adopting this GP term and rearranging \eqref{minmax}, we finally arrive at the following differentiable and fast-to-compute training objectives that approximately minimize the Wasserstein-1 distance defined in \eqref{inf}. The training objective for the D is 
\begin{align*}
    {\rm (P1.D)}~&\min_{\Theta_d}~\mathbb{E}\big[D(G(x_{\rm zf});\Theta_d)\big] - \mathbb{E}\big[D({y};\Theta_d)\big]  \\
        &\hspace{2cm} + \eta \mathbb{E}\big[(\| \nabla_{\hat{{x}}} D(\hat{{x}};\Theta_d)\| - 1)^2 \big]
\end{align*}
where $\Theta_d$ is the network parameters in D and $\eta$ controls the strength of the GP. The random sample $\hat{x}:=\alpha G(x_{\rm zf}) + (1-\alpha)y$ with $0 \leq \alpha \leq 1$.

The specific training objective for the G is derived directly from \eqref{minmax} as
\begin{align*}
{\rm (P1.G)}~~&\min_{\Theta_g}  -\mathbb{E}\big[D(G({x_{\rm zf}};\Theta_g))\big] 
\end{align*}
where $\Theta_g$ is the network parameters in G and $x_{\rm zf}=\Phi^{\dagger}(Y)$ is the zero-filled (ZF) image (inverse Fourier reconstruction from the ZF undersampled k-space measurements) input to the G. We refer to the above two equations as the unpaired WGAN objectives.

\vspace{1mm}

\noindent\textbf{Remark 1 [Enforcing 1-Lipschitz]}. Notice that there are more effective techniques for enforcing the 1-Lipschitz constraint in \eqref{minmax}. Two of those include spectral normalization \cite{spect} using singular value decomposition of weights, and computing $c$-transform over minibatches which implicitly enforces the constraints \cite{howwell}. These techniques can better satisfy the constraint at the expense of more computations and less expressive power of the D to estimate \textit{all} 1-Lipschitz functions.

\vspace{2mm}

\noindent\textbf{SGD algorithm}. 
$\Theta_g, \Theta_d$ are updated in an alternating fashion based on the SGD to optimize for (P1.D) and (P1.G) during training for each mini-batch of size $b$. First, the random samples $\{\hat{x}_i\}_{i=1}^b$ are drawn by uniformly sampling $b$ different $\alpha$s and linearly combining the corresponding G output and label in the current mini-batch. The mini-batch gradient of (P1.D) w.r.t. $\Theta_d$ is calculated given the labels $\{y_i\}_{i=1}^b$, the G outputs $\{x_i\}_{i=1}^b$, and random samples. Likewise, the G gradient (P1.G) is calculated, and the gradient steps are updated iteratively. 

\begin{figure}  [t]
\begin{center}
	\includegraphics[scale = 0.37] {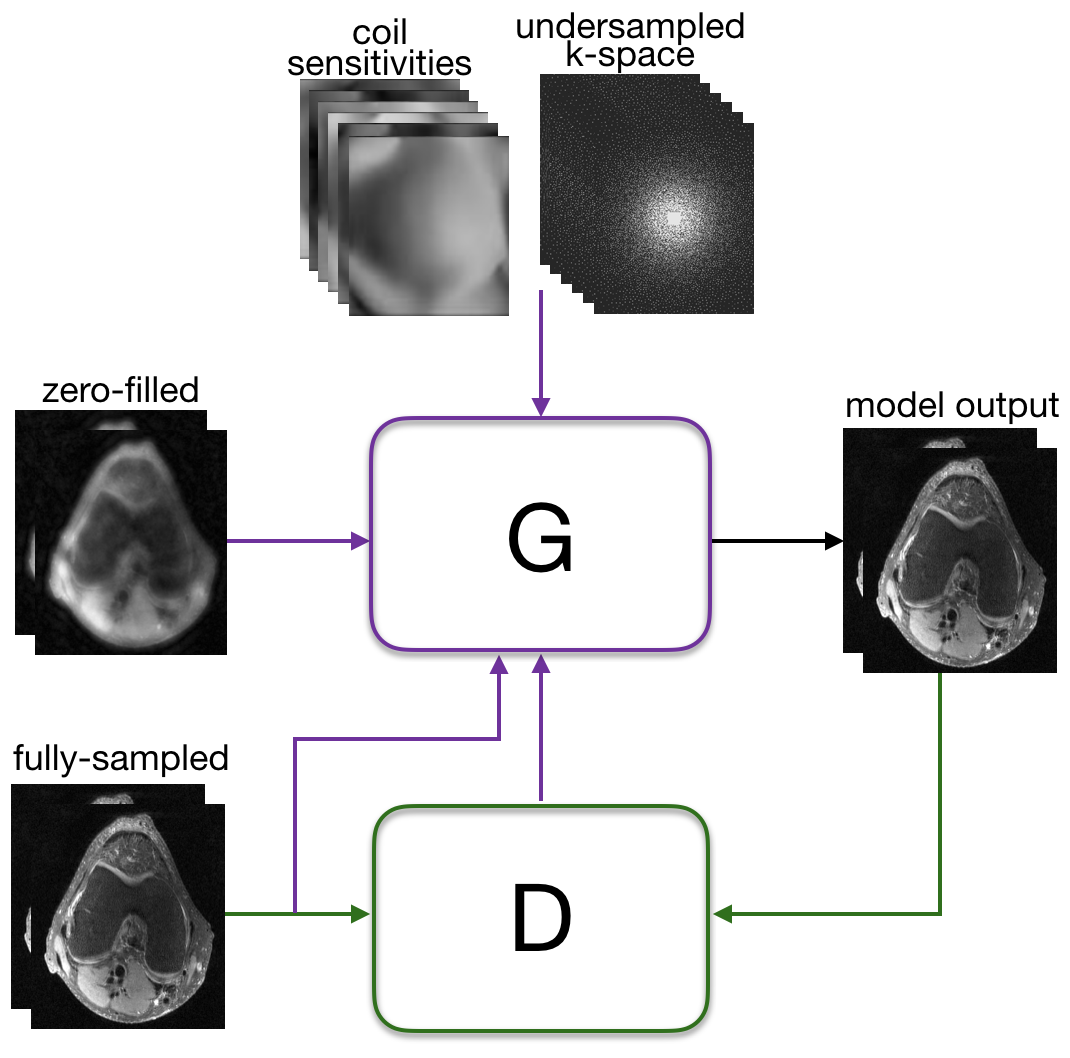}
    \caption{\small{Paired training with adversarial objectives.}}
    \label{flow_paired}
\end{center}
\end{figure}

\subsection{Paired training}
Supervised learning of the inverse mapping is common in the MR imaging context using pixel-wise losses. These approaches achieve stable training but the resulting images are typically blurry especially at high undersampling rates. \cite{GANCS} shows that adding adversarial training to the pixel-wise supervised training improves the sharpness and perceptual quality of the reconstructed images. LSGAN \cite{lsgan} objectives are combined with pixel-wise $\ell_1$ supervision in their work where the $\ell_1$ supervision helps control the high-frequency noise and stabilize their training. 

In our case, although the adversarial training alone is already relatively stable, adding more supervision when possible with a pixel-wise objective further improves the reconstruction quality. We find a pure $\ell_1$ objective gives superior results than a pure $\ell_2$ objective so $\ell_1$ is used for the pixel-wise supervision. Now G aims to output images close to its ground-truth label in terms of $\ell_1$ distance, and simultaneously gain a high score from D. The pixel-wise supervision is added to the G objective in (P1.G) which becomes 
    \vspace{-3mm}

\begin{equation}
\min_{\Theta_g}  -(1-\lambda)\mathbb{E}\big[D(G({x_{\rm zf}};\Theta_g))\big] + \lambda \mathbb{E}\big[\big\|{y}-G(x_{\rm zf};\Theta_g) \big\|_1\big]. 
\end{equation}
\vspace{1mm}

We consider two models when paired training is possible. When $\lambda<1$, the D is trained with the same objective defined by (P1.D), and we refer to the model as WGAN$+\ell_1$ hybrid model. We find that starting with $\lambda= 1$ and linearly increasing it with training steps provides a more refined initial phase and leads to a higher-quality final output. When $\lambda=1$, the training loss is the (paired) $\ell_1$ loss, only the G is involved, and we refer to the model as $\ell_1$-net. This is the traditional pixel-wise supervised paired training. Figure \ref{flow_paired} illustrates the paired training procedure of our hybrid model.

\begin{figure}  [!htbp]
\begin{center}
	\includegraphics[scale = 0.2] {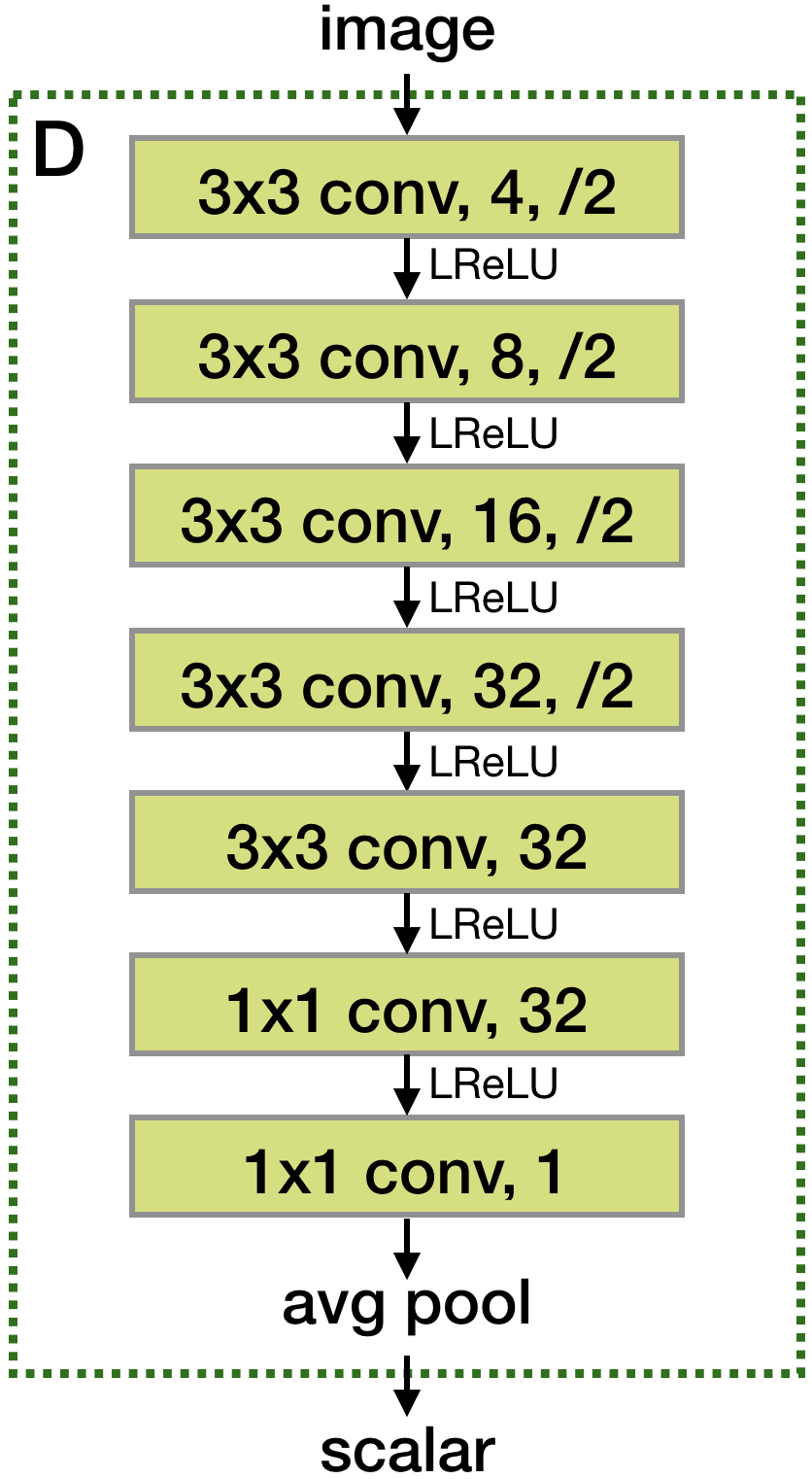}
	\includegraphics[scale = 0.325] {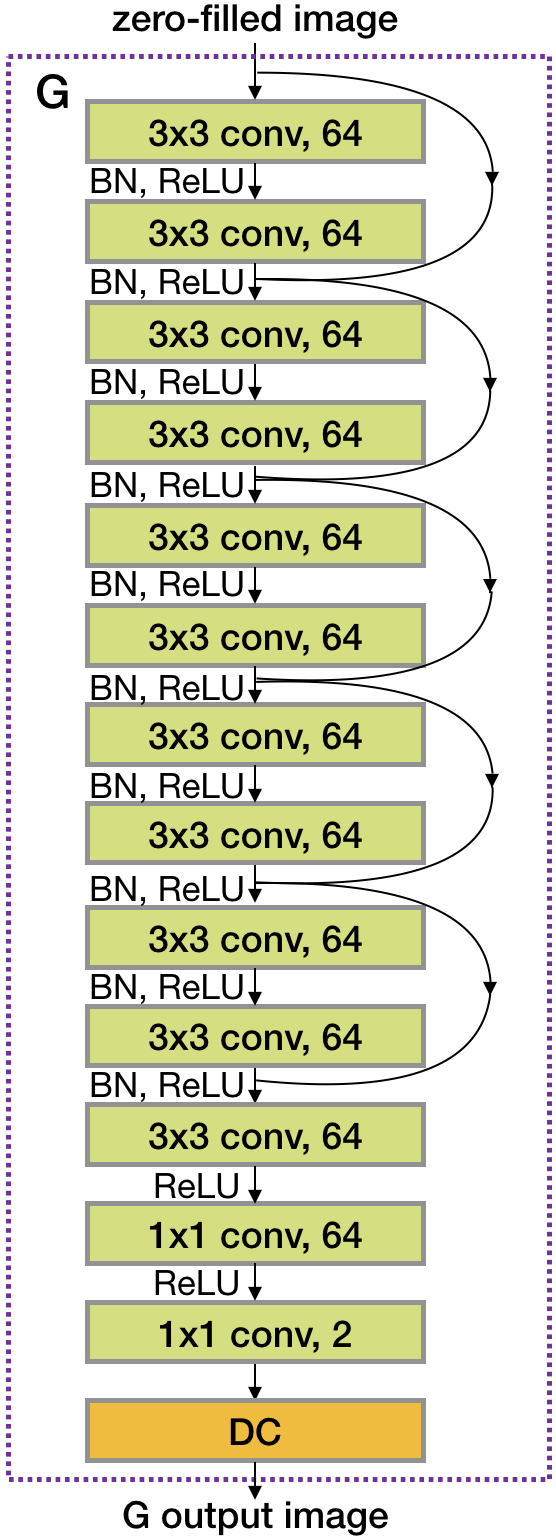}
    \caption{The discriminator network (left) and plain generator network with `hard' DC (right). BN and ReLU are applied after the summation with skip connections.}
    \label{plainGD}
\end{center}
\end{figure}
\vspace{-3mm}

\subsection{Generator networks with data consistency}
\begin{figure}  [!t]
\begin{center}
	\includegraphics[scale = 0.375] {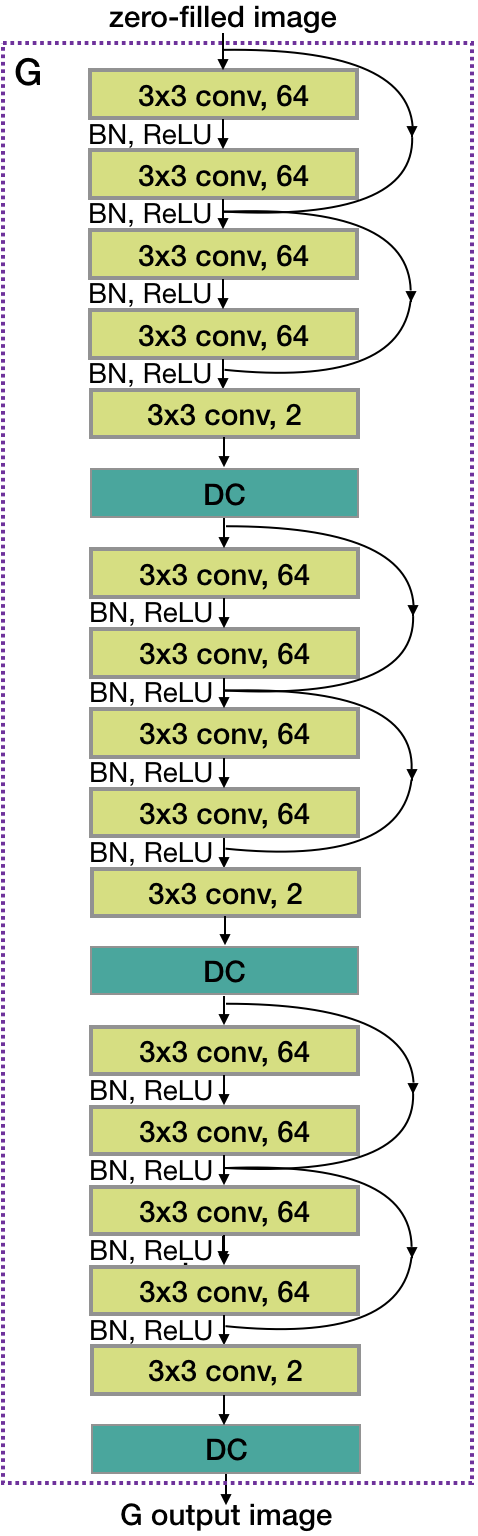}
    \caption{\small{The unrolled generator network with `soft' DC.}}
    \label{unrolled}
\end{center}
\end{figure}

The G network takes the zero-filled input image, $x_{\rm zf}$, which is simply an inverse DFT on the fully-sampled k-space masked (i.e. element-wise multiplied) by zeros and ones. Two channels are used to represent real and imaginary parts of an image separately. The G is supposed to output a high-quality version of its input image as visually close as possible to the labels' quality. Our training methods work with two types of G networks and data consistencies: from a standard ResNet \cite{ResNet} with a `hard' DC (Fig. \ref{plainGD}) (as that in \cite{GANCS}) to the state-of-the-art unrolled network with iterative `soft' DC (Fig. \ref{unrolled}). 

Unrolled networks were introduced recently and show superior performance for image recovery and restoration tasks \cite{abs-1805-03300,nips,DBLP:journals/corr/abs-1812-08115,Yang2018ADMMCSNetAD,bcd}. They are inspired by iterative inference algorithms \cite{SPIRiT}. The iterative process can be envisioned as a state-space model which at the $k$-th iteration takes an image estimate $x_k$, moves it towards the affine subspace of data consistent images, and then applies a proximal operator to obtain $x_{k+1}$. The state-space model is expressed as
\begin{align}
   &v_{k+1}=g(x_k) \\
   &x_{k+1} = NN(v_{k+1})
\end{align}
where $g$ is a DC operation with a learnable step size $\mu$ that combines the ZF data with the output of the previous iteration, $x_k$. Unfolding this recursion for a fixed number $K$ of iterations, one ends up with a recurrent NN (Fig.~\ref{unrolled}), where $x_K$ is the generator output.

The data consistency (DC) step ensures the k-space of the generated image is consistent with the actual input k-space data. `Soft' DC used in the unrolled network is a gradient descent step \cite{DBLP:journals/corr/DiamondSBWH17}
\begin{equation}
\label{gd}
    g(x)=x+\mu \ \Big[\sum_{i=1}^c  \mathcal{F}^{-1}\big\{\Omega\odot\mathcal{F}\{x\odot s_i\}\big\}\odot s_i^H-x_{\rm zf}\Big]
\end{equation}
where there are $c$ coil maps $s_i \in \mathbb{C}^{n}$. Alternatively, a simpler `hard' DC can be used at the end of a plain network with no need for learnable parameters:
\begin{equation}
\label{ndcm}
    g(x)=\sum_{i=1}^c  \mathcal{F}^{-1}\big\{Y^i+(1-\Omega)\odot\mathcal{F}\{x\odot \mathbf{s}_i\}\big\}\odot \mathbf{s}_i^H
\end{equation}
where $Y^i \in \mathbb{C}^{n}$ is the binary $\Omega$ masked k-space measurement from the $i$th coil.

\begin{figure*}[!htbp]
\begin{center}
\hspace{0.06cm} input (3-fold) \hspace{1.5cm} EGAN \hspace{1.6cm} LSGAN  \hspace{1.5cm}  WGAN \hspace{1.5cm} SOUP-DIL \hspace{1.4cm} fully-sampled

	\includegraphics[scale = 0.67] {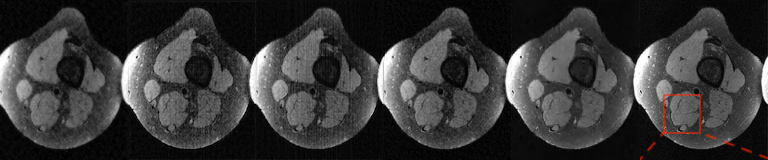}
		\includegraphics[scale = 1.19] {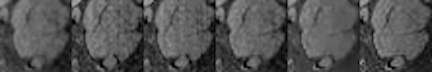}

    \caption{
    A representative test sample for unpaired training under different adversarial objectives, namely EGAN, LSGAN, WGAN-GP. For the single-coil acquisition model, we use a ResNet as the generator (plain G) model with 17 subjects for input and 6 subjects for labels. For completeness, we also compare with the SOUP-DIL scheme in \cite{soup} that models the image via sparse dictionary-learning. The bottom row shows a zoomed-in region.}
    \label{lsgan}
\end{center}
\end{figure*}

\subsection{Discriminator network}
D takes two kinds of inputs: the G output and the labels. For paired training, both inputs are complex-valued and represented by two channels. For unpaired training, D can also take single-channel magnitude images. We explore this relaxation so that datasets which consist of only magnitude images and no k-space data can also be used as labels. The G output is always complex-valued and is converted to a magnitude image before feeding to the D when the label is magnitude image. 

D outputs a real-valued scalar. A 7-layer plain CNN is used for D as shown in Fig. \ref{plainGD}, where the architectural details are provided. For the first four layers, the number of feature maps is doubled from 4 to 32, and a stride of 2 is used. Leaky ReLU nonlinearity (LReLU) \cite{maas2013rectifier} activation is used for all layers except the last one. The last layer averages out the seventh layer features to end up with a scalar score.

\section{Experiments}
The effectiveness of the unpaired WGAN scheme is assessed for single- and multi-coil MR acquisition models with Cartesian sampling. Our experiments and evaluations are performed to compare unpaired and paired deep learning models combined with five different training objectives. \textcolor{black}{To show the generalizability of the proposed unpaired training, we consider experimental settings that vary in four aspects: number of coils, model architecture, label availability, and undersampling ratio.} 

\noindent\textbf{Knee MRI dataset.} This dataset\footnote{Available at mridata.org} \cite{data} includes 19 subjects scanned with a 3T GE MR750 whole-body MR scanner. Each subject’s knee was placed in an 8-channel HD knee coil. Fully sampled images are acquired with a 3D FSE CUBE sequence with proton density weighting including fat saturation. Other parameters include FOV = 160 mm (sagittal), TR = 1550 ms, TE = 25 ms, slice thickness 0.6 mm (sagittal). For each subject we have a complex-valued 3D volume of size $320 \times 320\times 256$. The fully-sampled data used for reference images below takes over 41 minutes to collect for one subject. Axial slices of size $320\times 256$ are the input for training and test. For the partial label case (IV.B), 17 subjects are used for training and 2 subjects for testing. For the disjoint label and paired case (IV.C, D), 13 subjects are used for training and 6 subjects for testing. {In the partial label case we aim to find the minimum level of label availability, and more data is included in the training set to better differentiate the three levels being examined. In the disjoint label case, 4 more subjects are left to the test set for more confident conclusions and the reader study.} The inputs are undersampled by a variable density Poisson mask $\Omega$ with a fully-sampled center of size $20\times20$. 

\subsection{Network architecture and training}
The plain G network is a deep ResNet~\cite{ResNet} with $5$ residual blocks (RBs) followed by $3$ Conv layers. The D network consists of $7$ Conv layers with LReLU nonlinearity; see Fig.~\ref{plainGD}. Also, as shown in Fig.~\ref{unrolled}, the unrolled G has $K=3$ iterations, each with two RBs. Batch normalization (BN)~\cite{BNor} and ReLU are used after each layer except the last Conv layer for both plain and unrolled G. We set the gradient penalty coefficient $\eta=10$, which is the value suggested by the original WGAN-GP work \cite{WGAN_2}, and the value used by almost all work utilizing WGAN-GP objectives. Adam optimizer is used with the momentum parameter $\beta = 0.9$, mini-batch size 4, and learning rate $10^{-4}$. For paired GAN+$\ell_1$ training, ~$\lambda=0.99$ is used. Fully-sampled images are windowed to increase the brightness of the labels. The model is implemented in Tensorflow and the source code is available online at GitHub\footnote{https://github.com/lisakelei/Unpaired-GANCS/}.

\subsection{Unpaired training with partial labels} 
We start with a single-coil plain G model (as in the work GANCS \cite{GANCS}) for the partial labels scenario. Undersampled data are obtained by applying a $n$-fold undersampling mask to the `k-space' of the fully-sampled image. Fully-sampled k-space in the single-coil case is obtained by a 2D DFT of the complex-valued image reconstructed from the actual fully-sampled multi-coil measurement. The inputs to the single-coil model are 3-fold undersampled. 

We first show that WGANs is indeed more suitable for our task than EGANs\cite{GAN} and LSGANs\cite{lsgan}. Here we remove all $\ell_1$ supervision and use $17$ subjects as the inputs ($M=5440$) and $6$ subjects as the labels ($N=1920$). GANCS trained without $\ell_1$ objective, that is, with merely LSGANs or EGANs objective, outputs images with heavy coherent artifacts (Fig. \ref{lsgan}).

We also compare the unpaired methods to a blind sparsity penalized learning method, sum of outer products dictionary learning (SOUP-DIL) \cite{soup}, which does not require training labels. We use the $\ell_0$-norm penalized formulation (SOUP-DILLO)  which is the best performing variation in \cite{soup}. \textcolor{black}{Among its six scalar hyperparameters and one array hyperparamter, we tune the two most influential ones for the best PSNR.   The values we use for the array of sparsity penalty weights: $\lambda$ decreasing logarithmically from 0.3 to 0.01; for the patch size: $n=32$. The number of outer and inner iterations are by default set to 45 and 5, respectively.}
The inference time of this method is three orders of magnitude slower than our CNN based method (Table \ref{time}). Fig. \ref{lsgan} shows that outputs from this method tend to look smooth and blurry. Table \ref{table8} lists the \vspace{1mm} quantitative performance in terms of PSNR ($10\hspace{0.8mm} \text{log}_{10}\frac{\max_j |y_j|^2}{\hspace{0.5mm}\frac{1}{J}\hspace{-0.8mm}\sum_{j=1}^{J} |y_j-x_j|^2}$)\vspace{1.2mm} and SSIM of the four methods, averaged over 640 image slices from two test subjects. 

\begin{table} [!h]
  \caption{{Quantitative evaluations of unpaired training under different adversarial losses, compare with blind dictionary learning SOUP-DILLO.} } 
  \label{table8}
  \centering
  \begin{tabular}{|c||c|c|c|c|}
    \hline
    Method     &  EGAN  & LSGAN & WGAN-GP & SOUP-DILLO\cite{soup}\\
    \hline
    PSNR &  30.21  & 31.41 & 34.18  &33.53\\
    SSIM & 0.821 & 0.844 & 0.904 &0.886\\
    \hline
  \end{tabular}
\end{table}

\begin{table*} [!h]
  \caption{\textcolor{black}{Quantitative evaluations of the multi-coil plain model trained with unpaired GAN losses and paired $\ell_1$ loss. 17 subjects of 5 to 9-fold undersampled inputs, and 3 to 17 subjects of labels are used for training. Cases not specified with a loss type are from the proposed WGAN unpaired training.}}
  \label{snr}
  \centering
  \begin{tabular}{|c||c|c|c|c|c|c|c|c|c|c|}
    \hline
    Experiments & \multicolumn{3}{|l|}{\hspace{2.1cm} 5 fold} & \multicolumn{5}{|l|}{\hspace{3cm} 7 fold} & \multicolumn{2}{|l|}{\hspace{0.9cm} 9 fold} \\    
    \cline{2-11}
         &  $\ell_1$ 17 subjects  & 6 subjects & 3 subjects & $\ell_1$ 17 sub & EGAN 6 sub &LSGAN 6 sub & 6 sub & 3 sub & 6 subjects & 3 subjects\\
     \hline
    PSNR &  36.03  & 34.92&34.51& 35.86&29.62 &26.56 & 33.05 &32.84 &30.96& 30.23  \\
    SSIM & 0.847 & 0.873 & 0.869 & 0.811&0.657 &0.611& 0.842& 0.835 &0.819&0.813\\
    \hline
  \end{tabular}
\end{table*}

\begin{figure}  [!h]
\begin{center}
	input (3-fold) \hspace{0.1cm} 3 subjects \hspace{0.1cm} 6 subjects   \hspace{0.1cm} fully-sampled
	\includegraphics[scale = 0.5] {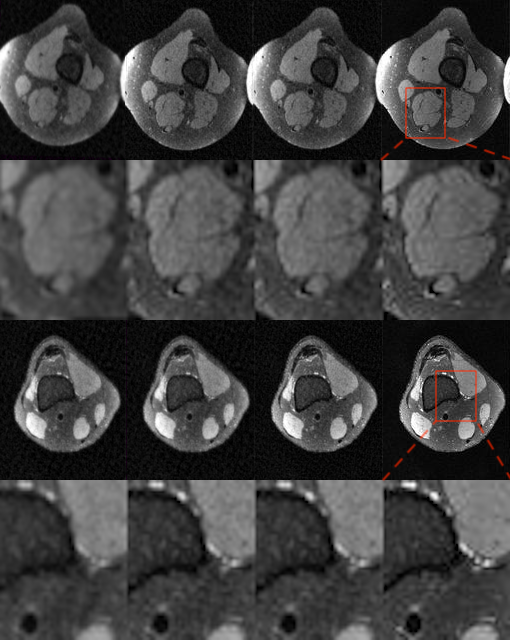}
    \caption{Two representative test samples from the single-coil model. From left to right: 3-fold undersampled input, output from our model trained with 3 subject and 6 subject labels, and fully-sampled reference.}
    \label{singleWGAN}
\end{center}
\end{figure}

\begin{figure*}  [!h]
\begin{center}
\vspace{3mm}
input (7-fold)\hspace{1.6cm}  EGAN \hspace{1.6cm} LSGAN \hspace{1.6cm} WGAN \hspace{1.6cm} $\ell_1$-net  \hspace{1.4cm}  fully-sampled 
\vspace{1mm}
	\includegraphics[scale = 0.21] {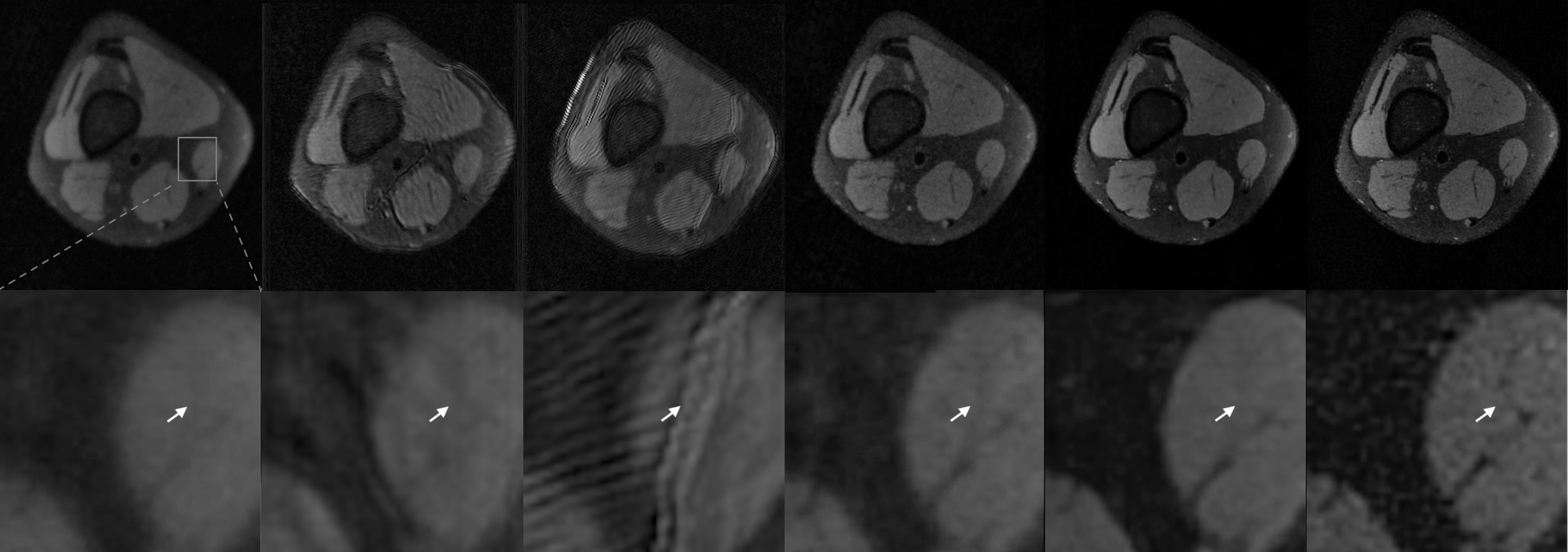}
    \caption{\small{A representative test sample from the multi-coil plain model trained with four different training losses. From left to right: 7-fold undersampled input, outputs from unpaired EGANs, LSGANs, and WGANs training with 6 subject labels, output from paired $\ell_1$ training with 17 subject labels, fully-sampled reference.}}
\label{7f6pl1}
\end{center}
\end{figure*}

We then test WGAN based unpaired training with different numbers of partially available labels. The single-coil model is trained using inputs of undersampled volumes from $17$ subjects ($M=5440$) and labels of fully-sampled volumes from $3$ and $6$ subjects ($N=960$ and $N=1920$). Two sample images are shown in Fig. \ref{singleWGAN}. Qualitatively, the result from $N=960$ is similar to that from $N=1920$.

All subsequent experiments are done with a multi-coil model with k-space data from 8 coils. The coil sensitivities are extracted by the ESPIRiT algorithm \cite{doi:10.1002/mrm.24751}.

\textcolor{black}{Again, we first show that WGANs is more suitable for our task than EGANs\cite{GAN} and LSGANs\cite{lsgan}. Here we use only adversarial objectives, $17$ 7-fold undersampled subjects as the inputs ($M=5440$) and $6$ subjects as the labels ($N=1920$). The training processes of EGANs and LSGANs did not converge. Therefore, we present in Fig. \ref{7f6pl1} the results from the epoch achieving the highest PSNR. These images are still heavily corrupted by coherent artifacts. The quantitative results over $640$ test slices are shown in Table \ref{snr}}.

Fig.~\ref{7f6pl1} also shows the outputs from unpaired WGANs training compared with that from paired $\ell_1$ training. Compared to a model trained with pixel-wise losses, our model trained with pure WGAN loss not only allows for using fewer labels but also generates images with more realistic texture. Pixel-wise paired training (with double the labels of the unpaired training) while refining the edges better, oversmooths images.

We examine the WGANs based unpaired training with the same $M$ and $N$'s but different undersampling ratios of 5, 7, and 9. Table \ref{snr} lists the average PSNR and SSIM over 640 slices from two test subjects.  Note that minimizing the paired $\ell_1$-loss encourages maximizing PSNR so $\ell_1$-net tends to get a higher PSNR regardless of its visual quality. Quality of the output images varies with the undersampling ratio (i.e. quality of the input), but outputs from 3-subject labels are only 0.4 dB lower on PSNR and 0.006 lower on SSIM, on average, compared to those from 6-subject labels. The above experiments with the single and multi-coil models show that we can decrease the number of labels to approximately 18\% of that used in the paired case. \textcolor{black}{From these experiments we observe that the proposed unpaired training is applicable for situations where only a small fraction of input and labels are paired (even for less than $20\%$).}

\begin{figure*}  [h]
\begin{center}
	input (10-fold) \hspace{1.5cm} unpaired \hspace{2cm} $\ell_1$ \hspace{1.5cm}  WGAN+$\ell_1$ \hspace{1.5cm}  CS-WV \hspace{1.2cm} fully-sampled
	
\vspace{1mm}

	\includegraphics[scale = 0.33] {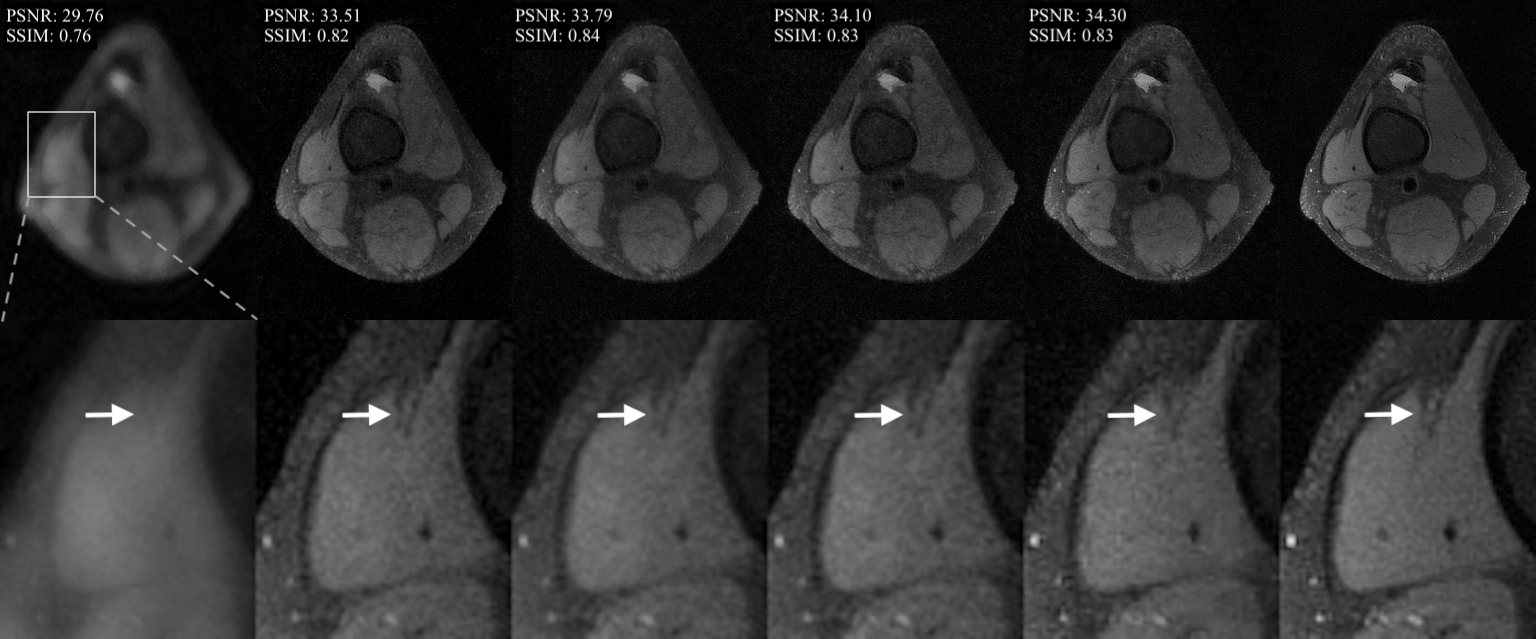}
    \caption{Test samples from the multi-coil unrolled model trained with three objectives compared with CS. From left to right: 10-fold input, outputs from unpaired model trained with WGAN, paired model trained with $\ell_1$, paired model trained with WGAN and $\ell_1$, CS-Wavelet, fully-sampled reference. }
    \label{10fall}
\end{center}
\end{figure*}

\begin{figure*}  [t]
\begin{center}
	input (10-fold) \hspace{1.5cm} unpaired \hspace{2cm} $\ell_1$ \hspace{1.5cm}  WGAN+$\ell_1$ \hspace{1.5cm}  CS-WV \hspace{1.2cm} fully-sampled
	
\vspace{1mm}

	\includegraphics[scale = 0.33] {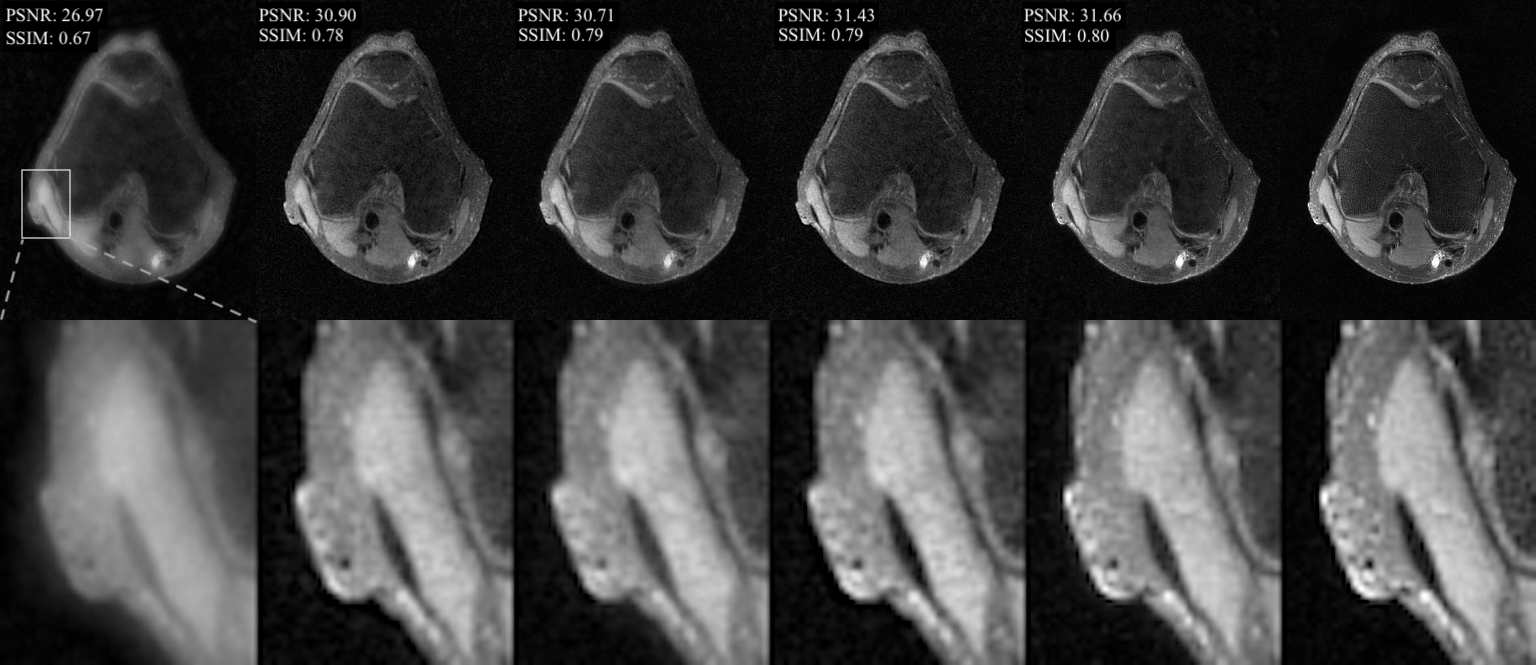}
    \caption{Another set of test samples from the multi-coil unrolled model trained with three objectives compared with CS. From left to right: 10-fold input, outputs from unpaired model trained with WGAN, paired model trained with $\ell_1$, paired model trained with WGAN and $\ell_1$, CS-Wavelet, fully-sampled reference.}
    \label{10fall07}
\end{center}
\end{figure*}

\begin{figure*}  [!ht]
\begin{center}
	input (6-fold) \hspace{2.4cm} unpaired \hspace{2.4cm}  CS-WV
	
\vspace{1mm}

	 \includegraphics[scale = 0.12] {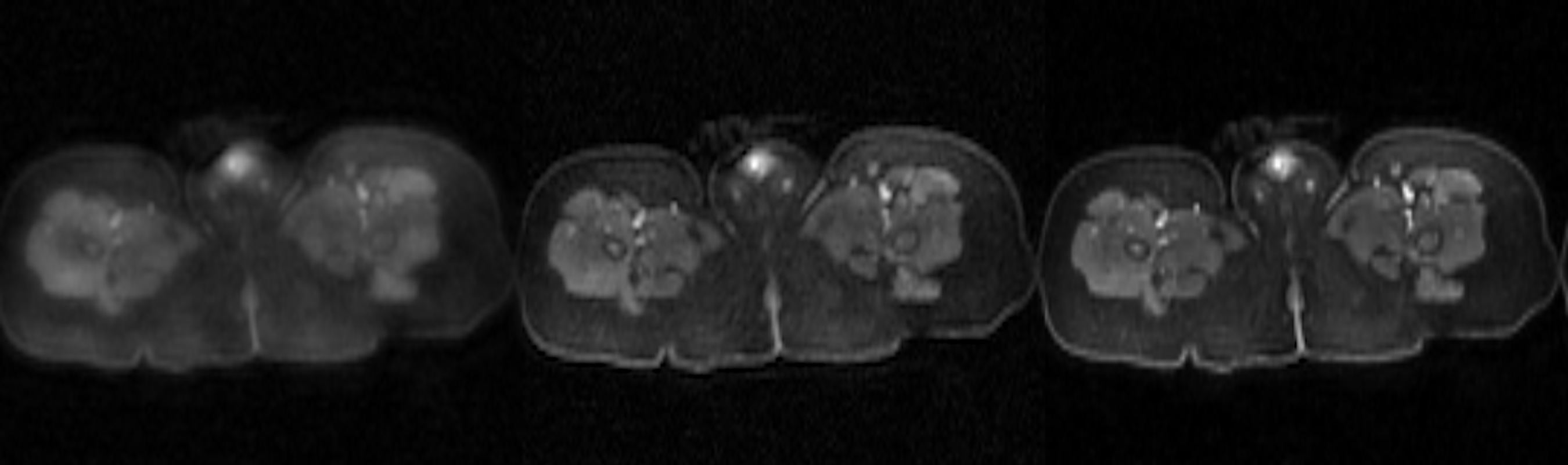}
	 \includegraphics[scale = 0.2813] {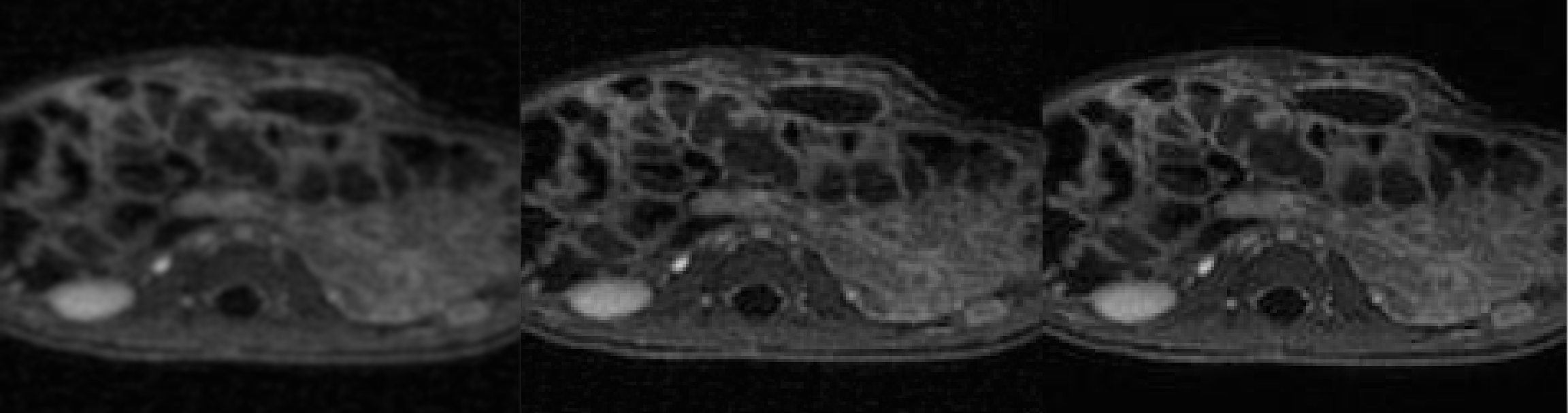}
    \caption{Two representative test samples of the multi-coil unrolled model (trained with disjoint knee datasets) and CS-Wavelet on DCE abdominal input.}
    \label{abd}
    \end{center}
\end{figure*}

\begin{table*} [!htbp]
  \caption{{Quantitative evaluations of the multi-coil unrolled model trained with three objectives compared with CS-WV.}}
  \label{table2}
  \centering
  \begin{tabular}{|c||c|c|c|c|c|}
    \hline
     \     &  input (10-fold)  & unpaired WGAN & $\ell_1$-net  & WGAN+$\ell_1$ & CS-Wavelet \\
    \hline
    PSNR &  28.23  & 32.73& 32.95& 33.25& 33.47  \\
    SSIM & 0.692 & 0.822 & 0.828 & 0.824& 0.831 \\
    \hline
  \end{tabular}
\end{table*}

\subsection{Unpaired training with disjoint datasets}
We now switch to the unrolled G which gives more accurate images (with around 2dB better SNR) compared to plain G and explore a more relaxed setting for the labels where there is no overlap between the input and label sets. Among the 13 subjects in the training set, undersampled raw data from 7 subjects are inputs, and fully-sampled magnitude images from 6 other subjects are labels. This setting reflects the case when we want to train a model for a dataset without any label using high-quality labels from some other datasets. 

We train the unrolled G with pure WGAN-GP \cite{WGAN_2} objective on 10-fold undersampled inputs. The inference sample and quantitative score from this model along with some other models are shown in Fig.\ref{10fall} and \ref{10fall07} and Table \ref{table2}. The quantitative scores are averaged over 1920 test slices, and only the center 272 x 216 region out of a 320 x 256 image is used.

The conventional CS method, which does not require training labels, can be used in the disjoint label setting thus included in the comparison. We use the CS-Wavelet implementation by the BART \cite{bart} toolbox. The regularization parameters $0.05$ is tuned to optimize the perceptual quality of a small evaluation set. Sample reconstructed slices are also shown in Fig.~\ref{10fall} and Fig.~\ref{10fall07}. 

{We also test the above unrolled G on a DCE abdominal dataset, for which fully-sampled data cannot be obtained. Fig. \ref{abd} gives two test samples from our unrolled model comparing with the CS-Wavelet reconstruction. It shows that our model can generalize well to different types of scan and anatomy even though the network is trained with unpaired knee data.}

Overall, the comparisons among these schemes and configurations indicate unrolled ResNets with WGAN training as the viable alternative to CS.

\subsection{Paired training}
In this section, we consider a supervised scenario with input and label pairs from 6 subjects. The network is trained with unrolled G and two different objectives.

We train the hybrid WGAN+$\ell_1$ model with the first 500 batches with pure $\ell_1$ objective, then linearly decrease $\lambda$ to 0.99 within the first 1000 mini-batches. This is useful to stabilize training and improve final performance. We also train a model with only $\ell_1$ objective ($\ell_1$-net). Two test slices from these two models are shown in Fig.\ref{10fall} and \ref{10fall07}, and the quantitative scores are shown in Table \ref{table2}. The conclusion is that when paired training is possible, adding WGAN objective to the classic $\ell_1$-minimization leads to results that are visually sharper with higher SNR. 

\begin{figure*}  [!h]
\begin{center}
	\includegraphics[scale = 0.385] {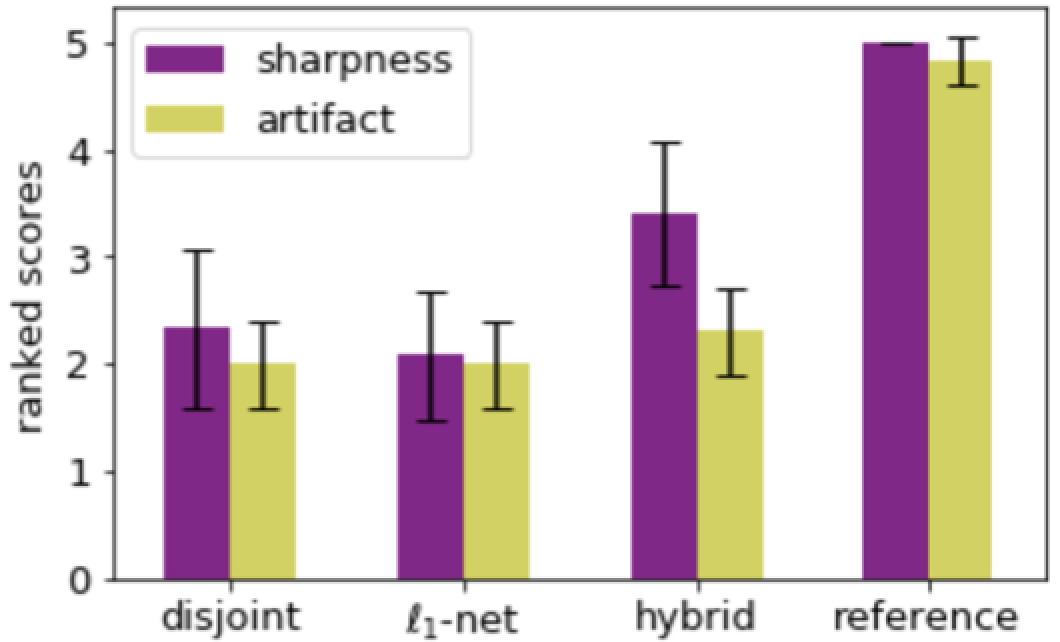}
	\hspace{7mm}
	\includegraphics[scale = 0.385] {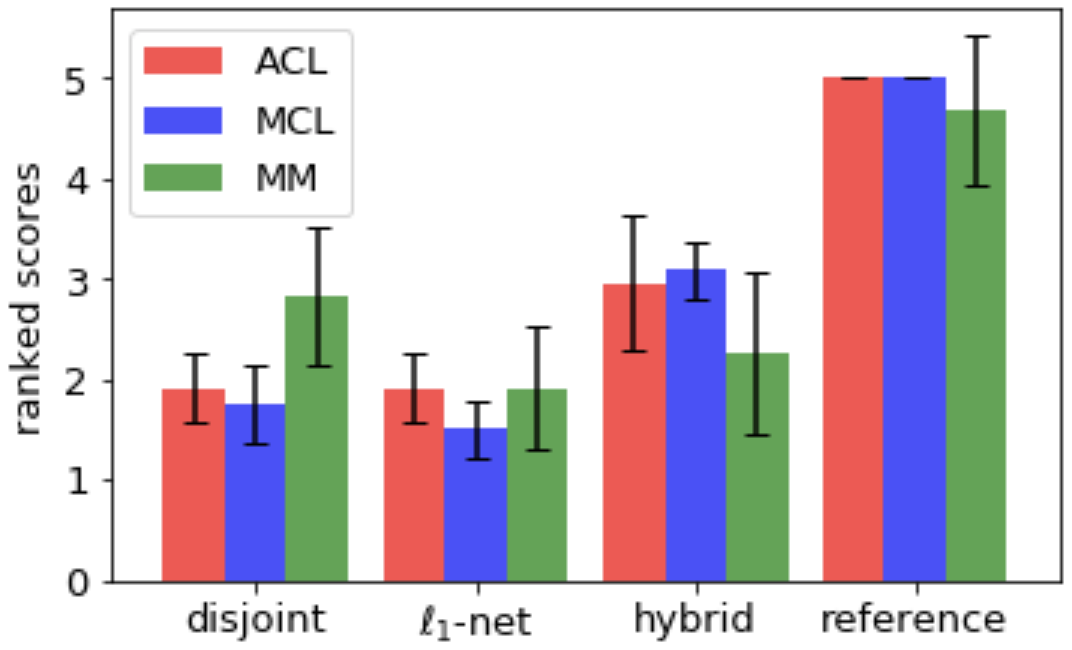}
    \caption{Ranked score from radiologist review for outputs from disjoint unrolled model, paired unrolled $\ell_1$ model, paired unrolled WGAN+$\ell_1$ (hybrid) model, and fully-sampled reference. Aspects of rating: sharpness of the image, level of coherent artifacts, and visibility of three knee structures: ACL, MCL, MM.}
    \label{ros}
\end{center}
\end{figure*}

\subsection{Radiologist evaluation}
We notice the standard quantitative metrics (i.e. SNR, PSNR, MSE, SSIM, etc.) including those reported above do not reflect the visual quality of the reconstructed images well. To assess the diagnostic image qualities from different reconstruction methods, we perform an experiment based on the consensus of two radiologists. We asked them to rank the reconstructed volumes given by four reconstruction methods together with the fully-sampled volume according to five aspects: sharpness, level of coherent artifacts, visibility of anterior cruciate ligament (ACL), medial meniscus (MM) and medial collateral ligament (MCL). ACL, MM, and MCL are three structures in the knee that are commonly assessed. 

Image volumes from 6 different subjects, 5 versions each, (30 volumes in total) are used for this test. Radiologists were blinded to the reconstruction schemes. Horos \cite{horos} software interface is used to visualize the images. For each subject, the five volumes (for different reconstructions) are ranked from best to worst with ties possible. We then convert the rankings to scores; the best score is 5, and the worst one is 1. When there is a tie, we take an average of the scores; for example, if the second and third best scores are equally good, both would receive the score $\frac{4+3}{2}=3.5$. The scores for all NN based methods are presented in Fig. \ref{ros}.

\begin{table*} [!htbp]
  \caption{{Inference time per slice of plain G,  unrolled G, CS-Wavelet, and SOUP-DIL.}}
  \label{time}
  \centering
  \begin{tabular}{|p{2cm}||p{2cm}|p{2cm}|p{3cm}|p{3.3cm}|}
    \hline
    Method     &  plain G  & unrolled G & 30-iter CS & SOUP-DIL \\
    \hline

    Time (sec)  &  0.022  & 0.025 & 0.563 & 196  \\
    \hline
    Implementation &\multicolumn{2}{|l|}{TITAN Xp GPU, TensorFlow}  & TITAN Xp GPU, C & Intel Xeon CPU, MATLAB\\
    \hline
  \end{tabular}
\end{table*}

\subsection{Inference time }
{Table~\ref{time} shows the average reconstruction time per 2D slice for our plain and unrolled generator models, CS, and SOUP-DIL~\cite{soup}. The timing starts after the initial data reading and ends before the final data writing. It is also averaged across two test volumes. Our methods, CS, and SOUP-DIL reconstruct 4, 8, and 1 slice at a time, respectively.     Our methods and the CS-Wavelet reconstruction using BART~\cite{bart} are implemented in TensorFlow and C, respectively. Both run on an NVIDIA TITAN Xp GPU. The official implementation~\cite{soupgit} of SOUP-DIL is in MATLAB and 
we run on an Intel Xeon Gold 6126 CPU.  The 3-iteration unrolled G is only slightly slower than the plain G. Both of our models are about 23 times faster than the conventional CS-Wavelet that takes advantage of a very efficient implementation using BART~\cite{bart}. Under the implementation settings shown in Table~\ref{time}, SOUP-DIL is about 9,000 times slower than our methods. Note that MATLAB is generally less computational efficient, but implementing SOUP-DIL on a GPU will not substantially decrease its inference time because the algorithm has nested iterative steps which are difficult to parallelize. }

\section{Conclusions}
This paper advocates an unpaired deep learning scheme for MRI reconstruction when high-quality training labels are scarce. Leveraging Wasserstein GANs with gradient penalty, a generator network based on plain or unrolled ResNets maps linear image estimates to mimic the image label distribution. The discriminator network then plays the role of a critic that estimates the distance of generator output images from the label images. The unpaired training objectives alleviates the need for pairing among the undersampled input and the high-quality labels. Our work far extends the scope of prior work \cite{GANCS} for imaging scenarios with scarce training labels and more realistic multi-coil models. Our experiments on knee and abdominal MRI datasets -- deploying two network architectures under different data configurations and training schemes -- corroborate the efficacy of Wasserstein distance based adversarial, and most importantly, unpaired, training with DC to give a faithful reconstruction of MRIs and is a viable alternative to slow conventional methods.   

In particular, the proposed unpaired training works with 18$\%$ of the labels needed in the paired case, and when the training input and label are from disjoint datasets. When pairing is possible, training an unrolled network with WGAN+$\ell_1$ objective is better than with either the WGAN or $\ell_1$ objective alone. And it is in some aspects better than CS-Wavelet reconstruction. All of our NN based models are 23 times faster than CS-Wavelet reconstruction and substantially faster than a dictionary learning method.

\bstctlcite{IEEEexample:BSTcontrol}
\bibliographystyle{IEEEtran}
\bibliography{ref}

\end{document}